\documentclass{emulateapj}
\usepackage{graphicx}
\begin{document}
   \title{{\ } Ion pickup by finite amplitude parallel propagating
Alfv\'en waves}
   \author{Xing Li$^1$, Quanming Lu$^2$ and Bo Li$^1$
          }
   \affil{1. Institute of Math. and Phys. Sciences, University of Wales, Aberystwyth,
   UK \\
             \email{xxl@aber.ac.uk}
              2. University of Science \& Technology of China, Hefei, Anhui, China}
    \begin{abstract}
Two scenarios of possible ion heating due to finite amplitude
parallel propagating Alfv\'en waves in the solar atmosphere are
investigated using a 1D test particle approach. 1. A finite
amplitude Alfv\'en wave is instantly introduced into a plasma (or
equivalently, new ions are instantly created). 2. New ions are
constantly created. In both scenarios, ions will be picked up by the
Alfv\'en wave. In case 1, the wave scatters ions in the transverse
direction leading to a randomization (or heating) process. This
process is complete when a phase shift of $\pm \pi$ in the ion
gyrospeed is produced between particles with characteristic parallel
thermal speed and particles with zero parallel speed. This
corresponds to $t={\pi \over k v_{th}}$ ($k$ is the wavenumber and
$v_{th}$ is the ion thermal speed). A ring velocity distribution can
be produced for a large wave amplitude. The process yields a
mass-proportional heating in the transverse direction, a temperature
anisotropy and a bulk flow along the background magnetic field. In
case 2, continuous ion creation represents a continuing phase shift
in the ion gyrospeed leading to heating. New particles are picked up
by the Alfv\'en wave within one ion gyroperiod. It is speculated
that the mechanism may operate in the chromosphere and active
regions where transient events may generate finite amplitude
Alfv\'en waves. To appear in ApJ Letters May 10, 2007.
\end{abstract}

\section{INTRODUCTION}

It has been widely speculated that the energy that heats the corona
comes from the convective flows in the photosphere. The energy is
somehow transported into the coronal part through the magnetic
field. It is natural to think that Alfv\'en waves channel the energy
to the corona. Indeed, these waves have been observed in the solar
atmosphere (Ulrich 1996), and are ubiquitous in the extended corona
--- the fast solar wind (Smith et al. 1995). However Alfv\'en
waves are difficult to dissipate in collisionless plasmas. Hence,
non-linear processes have been assumed to cascade the wave energy
from low to high frequencies where wave dissipation is readily
possible (Hollweg 1986, Li and Habbal 2003).

When Alfv\'en waves propagate in a partially ionized plasma,
neutral-ion collisions produce a channel for the  wave dissipation
(De Pontieu \& Haerendel 1998; De Pontieu et al. 2001; Leake et al.
2005). Ions and neutrals can be collisionally coupled. A slippage
between the ion and neutral populations leads to the wave
dissipation. If we consider neutrals are constantly ionized at the
chromosphere, we will show that these newly created He$^{+1}$ ions
will be picked-up by the wave and will be energized.

In this Letter, a new scenario, ion pickup by an Alfv\'en wave, is
explored. The pickup process can lead to the heating of ions, and it
must also dissipate the wave. In section 2, we discuss the ion
pickup process by an instantly introduced wave (or identically, the
ions are created instantly). In section 3, the pickup of
continuously created ions by an Alfv\'en wave is investigated.
Finally, in section 4 we discuss possible applications of the ion
pickup process in the solar atmosphere.

\section{Ion pickup by Alf\'en waves}

Consider a parallel propagating monochromatic dispersionless
Alfv\'en wave with angular frequency $\omega$ and wave-number $k$,
and $\omega=kv_A$($v_A$ is the Alfv\'en speed). The wave
electromagnetic field $\delta {\mathbf B}_w$ and $\delta {\mathbf
E}_w $ are
\begin{equation} \left\{ \begin{array}{l}
\delta {\mathbf B}_w=B_k [\cos \phi_k {\mathbf i}_x -\sin \phi_k
{\mathbf i}_y], \\\delta {\mathbf E}_w= -v_A {\mathbf B}_0/B_0
\times \delta {\mathbf B}_w, \end{array} \right.
\end{equation} ${\mathbf i}_x$ and ${\mathbf i}_y$ are unit vectors,
$B_0$ the background magnetic field, and $\phi_k=k(v_A t-z) $
denotes the wave phase. The motion of a particle is described by
\begin{equation}
m_j {d {\mathbf v} \over dt }= e_j [ \delta {\mathbf E}_w + {\mathbf
v} \times (B_0 {\mathbf i}_z +\delta {\mathbf B}_w)], {d z \over
dt}= v_z. \end{equation} \label{Q1}
 Let $u_\perp=v_x+i v_y$,
$v_\parallel=v_z$, and $\delta B_w=B_k e^{-i\phi_k}$, we have
\begin{equation}
{d u_\perp \over dt } + i \Omega_0 u_\perp =i(v_\parallel -v_A)
\Omega_k e^{-i \phi_k}, \label{Q2}
\end{equation}
\begin{equation}
{d v_\parallel \over dt} =-\mbox{Im}(u_\perp \Omega_k e^{i\phi_k}),~
{d z \over dt}=v_\parallel, \label{Q3}
\end{equation}
where $\Omega_0=e_j B_0/m_j$, $\Omega_k=e_j B_k /m_j$, and $j$
refers physical quantities of ion species $j$. As a first order
approximation, $v_\parallel\approx v_\parallel(0)$ is a constant,
where $v_\parallel(0)$ is the initial ion parallel velocity. The
approximation is valid when $\Omega_k/\Omega_0=B_k/B_0$ is small and
the wave frequency is low so $|\Omega_0| \gg |k(v_\parallel-v_A)|$.
With the initial condition $u_\perp=u_\perp(0)$ and $z=z(0)$, the
solution of Eq. (\ref{Q2}) for a low beta plasma is (Wu et al. 1997,
Lu et al. 2006)
 \begin{equation}
u_\perp=\left[u_\perp(0)+v_A{B_k \over B_0}e^{ikz(0)}\right]e^{-i
\Omega_0 t} -v_A{B_k \over B_0} e^{-ik[v_At-z]}. \label{Q4}
\end{equation}
where $z=z(0)+v_\parallel(0)t$ and we have adopted approximations
$\Omega_0-k[v_A-v_\parallel(0)]\approx \Omega_0$,
 $v_A-v_\parallel(0)\approx v_A$.
The last term of $u_\perp$ is the perturbed ion velocity of an
Alfv\'en wave. The first term is due to the gyromotion of the
particle and the modification of the gyromotion due to the wave.
Note, if $u_\perp(0)=-v_A (B_k/B_0)e^{-ikz(0)}$ and
$v_\parallel(0)=0$, the particle is surfing on the wave, there is no
gyromotion.

Now let us consider an ensemble of newly created particles with a
Maxwellian distribution. The particles' average parallel speed is
zero but the transverse component is $u_{\perp f}$
\begin{equation}
u_\perp(0)=u_{\perp r}(0)+u_{\perp f}(0)=u_{\perp r}(0) - \alpha v_A
{B_k \over B_0} e^{ikz}, \label{Q5}
\end{equation}
where $\alpha$, a constant, describes the degree that the newly
created particles are settled in the wave field. If $\alpha=1$, the
newly created particles are already settled.
 $u_{\perp r}$ denotes the random
perpendicular thermal speed. After time $t$ particles with different
initial position $z(0)=z-v_\parallel(0)t$ will arrive at $z$. The
average transverse velocity at $z$ is
$$U_\perp=-v_A {B_k \over B_0} e^{-ik(v_At-z)}+~~~~~~~~~~~~~~~
$$

$${1-\alpha \over \sqrt{\pi}v_{th}}\int^\infty_{-\infty} v_A {B_k
\over B_0} e^{ik[z-v_\parallel(0)t]} e^{-i \Omega_0 t}
e^{-\left({v_\parallel(0) \over v_{th}} \right)^2 }dv_\parallel(0)$$

 \begin{equation} =-v_A {B_k \over
B_0} e^{-ik(v_At-z)}+(1-\alpha)A_k v_A {B_k \over B_0} e^{ikz}
e^{-i\Omega_0t}, \label{Q6}
\end{equation}
where $A_k={1 \over \sqrt{\pi}}\int^\infty_{-\infty}\cos (kv_{th} t
x) e^{-x^2} dx=e^{-k^2v^2_{th}t^2 \over 4}$, $v_{th}=\left({2k_B
T_{j0} \over m_j} \right)^{1/2}$ , and $T_{j0}$ is the initial
temperature of species $j$. Eq (\ref{Q6}) illustrates the pickup of
ions by the wave. Note $A_k\approx 0 $ when $kv_{th}t \geq \pi$.
Hence, regardless of the value of $\alpha$, these particles will be
picked up by the Alfv\'en wave: eventually $U_\perp$ is determined
by the wave only. Subtracting (\ref{Q6}) from (\ref{Q4}), one finds
the random perpendicular velocity at $z$:
$$
u_\perp-U_\perp=u_{\perp r}(0) e^{-i\Omega_0 t} +(1-\alpha)v_A {B_k
\over B_0} e^{ik[z-v_\parallel(0)t]} e^{-i\Omega_0 t} $$
\begin{equation} -(1-\alpha) A_kv_A {B_k \over B_0} e^{i(kz-\Omega_0
t)}.  \label{Q7}
\end{equation}
Eq. (8) describes the particles' gyrospeed in the frame of
$U_\perp$. When $\alpha=0$ (initially the newly created particles
have zero flow speed), particles will be strongly scattered in the
phase space. The second term on the right hand side is crucial for
our understanding the heating in the pickup process. In the phase
space frame of $U_\perp$, a randomization process is complete when
particles with characteristic speed $v_\parallel(0)=\pm v_{th}$
moving from $z(0)=z-v_\parallel(0)t$ to $z$ are scattered in the
perpendicular direction and a phase shift $\pm \pi$ in the gyrospeed
relative to particles with $v_\parallel(0)=0$ is produced. This
translates to $kv_{th}t \approx \pi$. For a finite amplitude
Alfv\'en wave, if $v_A B_k/B_0>v_{th}$, the distribution function
will be a ring. Hence particles are strongly heated. On the other
hand, if $\alpha=1$, initially particles are already ``picked up" by
(surfing on) the wave. Only the initial random motion survives, and
there is no heating. One notices that if $v_{th}=0$, $A_k=1$, there
is also no heating. Hence, the heating is a warm plasma effect. With
$U_\perp$, the perpendicular temperature can be found
$$T_{\perp j} ={m_j \over
2K_B v_{th} \sqrt{\pi}} \int^\infty_{-\infty} |u_\perp-U_\perp|^2
e^{-\left({v_\parallel(0) \over v_{th}} \right)^2} dv_\parallel(0)
$$
\begin{equation}
=T_{j0}\left[ 1+{m_j B^2_k \over m_p \beta_j
B^2_0}(1-A^2_k)(1-\alpha)^2\right], \label{Q8}
\end{equation}
where $\beta_j=8\pi n_e k_B T_{j0}/B^2_0$ is the plasma beta of
species $j$. The heating is obviously mass-proportional.

Substituting Eq. (\ref{Q4}) into Eq. (\ref{Q3}), and letting
$v_A-v_\parallel(0)\approx v_A$ and $\Omega_0-k(v_A-v_\parallel(0)]
\approx\Omega_0$, one finds
\begin{equation}
v_\parallel=v_\parallel(0)+v_A(1-\alpha){B^2_k \over B^2_0}
\{1-\cos[\Omega_0t -kv_At -kv_\parallel(0)t]\}, \label{Q9}
\end{equation}
Using the same procedure to obtain $U_\perp$ and $T_\perp$, we find
 the average parallel velocity and temperature:
\begin{equation}
U_\parallel=v_A{B^2_k \over B^2_0}(1-\alpha)[1-A_k \cos(\Omega_0
t-kv_A t)], \label{Q10}
\end{equation}
$$
T_{\parallel j}=T_{j0}\left[1+{2m_j B^4_k (1-\alpha)^2 \over m_p
\beta_j B^2_0}[(C_k-A^2_k)\times \right. ~~~~~
$$
\begin{equation}\left.
\cos^2(\Omega_0t-kv_At) +(1-C_k)\sin^2(\Omega_0t-kv_At)]\right],
\label{Q11}
\end{equation}
where $C_k={1 \over \sqrt{\pi}}\int^\infty_{-\infty}\cos^2(kv_{th}
tx)e^{-x^2}dx=0.5+0.5e^{-k^2v^2_{th}t^2}$. When $t\rightarrow
\infty$, $A_k=0$, and $C_k=0.5$.
We then obtain asymptotic values:
\begin{equation}
U_\perp=-v_A {B_k \over B_0}
e^{-ik(v_At-z)},~~U_\parallel=v_A(1-\alpha){B^2_k \over B^2_0}
\label{Q12}
\end{equation}
$$T_{\perp j}=T_{j0}\left[1+{(1-\alpha)^2 m_j B^2_k
\over m_p \beta_j B^2_0} \right],
$$
\begin{equation}
T_{\parallel j}=T_{j0}\left[1+{m_j B^4_k(1-\alpha)^2 \over m_p
\beta_j B^4_0} \right] \label{Q13}.
\end{equation}

\begin{figure}[b]
\epsscale{0.95} \plotone{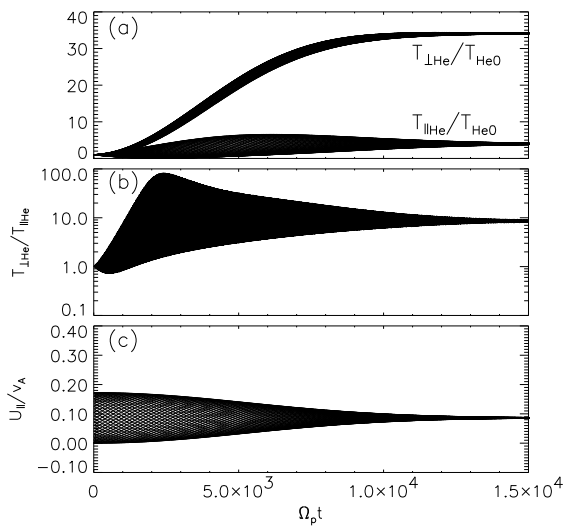} \caption{The time history of (a)
the perpendicular temperature $T_{\perp He}/T_{He0}$ and the
parallel temperature $T_{\parallel He}/T_{He0}$, (b) the temperature
anisotropy $T_{\perp He}/T_{\parallel He}$, and (c) the average
parallel velocity $U_\parallel/v_A$. Here we choose the particle as
He$^{+1}$, and the parameters are: $\alpha=0$,
$\omega/\Omega_p=0.25\pi/125\approx 0.0063$, $\beta_{He}=0.01$,
$B^2_k/B^2_0=0.09$. }
\end{figure}

\begin{figure*}[t]
\epsscale{0.84} \begin{center}\plotone{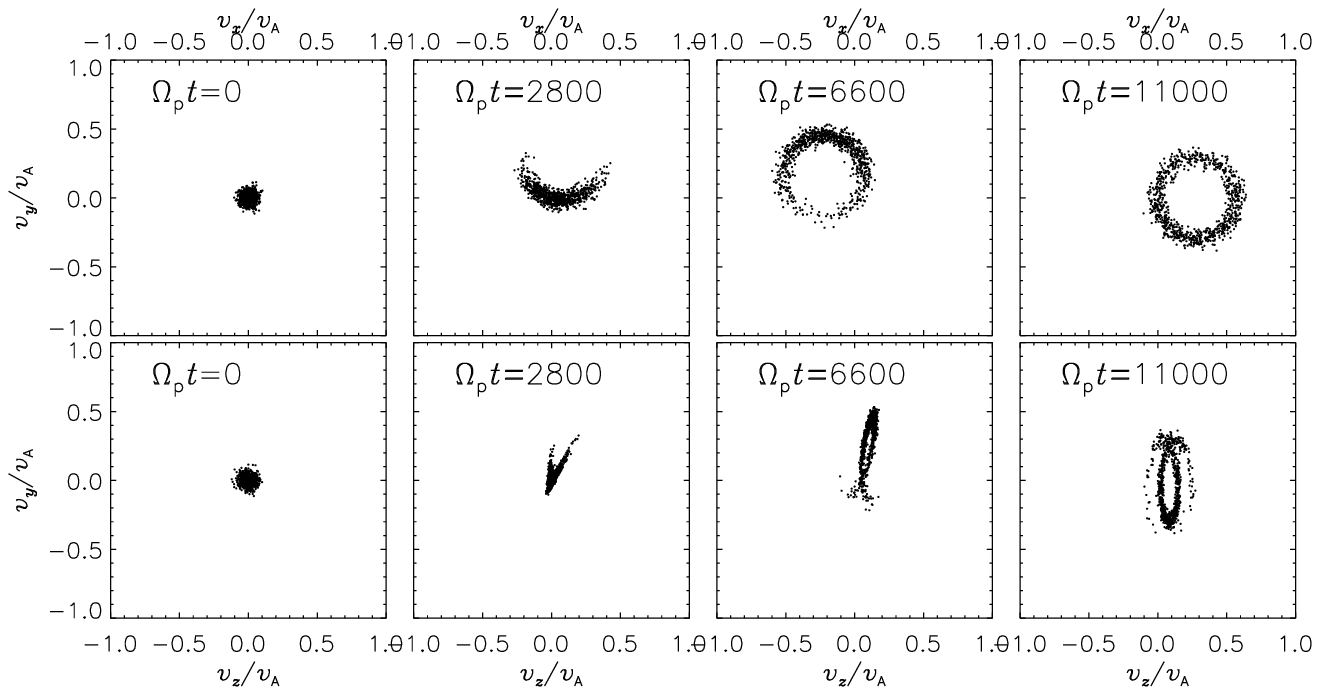}\end{center}
\caption{Top panel: Scatter plots of helium He$^{+1}$ velocity in
the velocity plane perpendicular to $v_\parallel$. Bottom panel:
scatter plots of helium He$^{+1}$ velocity in the plane parallel to
the background magnetic field. Relevant parameters are those in
Fig.1.}
\end{figure*}
%It is very interesting that a double ring velocity distribution can
%be found if the wave amplitude is sufficiently large.
Subtracting
(\ref{Q10}) from (\ref{Q9}), when $t\rightarrow \infty$, one finds
\begin{equation}
v_\parallel-U_\parallel=v_\parallel(0) +v_A(1-\alpha){B^2_k \over
B^2_0} \cos [\Omega_0 t -kv_At -kv_\parallel(0)t].  \label{Q14}
\end{equation}
It is possible that the second term on the right hand side is larger
than the initial random thermal velocity $v_\parallel(0)$. In this
case, the velocity distribution plotted in the $v_\parallel -v_y$
plane may be a ring as well (see Fig.2).

To verify the above analysis, a test particle simulation was
conducted by calculating the full dynamics of particles (Eq.
\ref{Q1}) in the electromagnetic field of a given Alfv\'en wave
described by Eq. 1. The equations were solved using the Boris
algorithm with time step $\Delta t=0.01\Omega^{-1}_p$ ($\Omega_p$ is
the proton gyrofrequency) and periodic boundary conditions.
Initially, 200000 particles with Maxwellian velocity distribution
were evenly distributed in a region with length
$1000v_A\Omega^{-1}_p$ (1000 cells). The flow speed of these
particles was zero. The average parallel velocity, and the parallel
and perpendicular temperatures were obtained by using the following
procedure: we first computed $U_\parallel=\left<v_z\right>$,
$T_\parallel={m_j \over k_B}\left<(v_z-\left<v_z\right>)^2\right>$,
and $T_\perp={m_j \over
2k_B}\left<(v_x-\left<v_x\right>)^2+(v_y-\left<v_y\right>)^2\right>$
in every cell ($\left<\right>$ denotes an average over a cell).
These quantities were then averaged over all cells, so only random
motions contribute to the temperatures. Fig.1 shows the time
evolution of (a) the perpendicular temperature $T_{\parallel
He}/T_{He0}$ (here 0 refers to the initial value) and the parallel
temperature $T_{\perp He}/T_{He0}$, (b) the temperature anisotropy
$T_{\perp}/T_\parallel$, and (c) the average parallel velocity
$U_\parallel/v_A$ of the test particle simulation. The ``shaded
areas" are due to rapid ion gyrations. The results are consistent
with our analytical predictions. At $\Omega_pt \approx 12000$, an
asymptotic stage is reached: $U_\parallel/v_A \approx 0.086$,
$T_{\perp He}/T_{He0}\approx 34$ and $T_{\parallel
He}/T_{He0}\approx 3.9$ with the temperature anisotropy $T_{\perp
He}/T_{\parallel He}\approx 8.7$.

Fig.2 displays scatter plots of the helium ions in the five cells
around $z=500v_A\Omega^{-1}_p$  at $\Omega_pt=0$, 2800, 6600 and
11000. Initially, helium ions satisfy the Maxwellian distribution
with $v_{th}=0.005v_A$ and $\beta_{He}=0.01$. Helium ions are
dramatically scattered in the transverse direction in the phase
space. The ring velocity distributions are nicely shown in Fig.2.
The ring velocity distribution is unstable (Lu and Wang 2005). The
velocity distribution may eventually be thermalized by relevant
microscopic instabilities of the ring distribution. The bulk
acceleration of particles along ${\mathbf B}_0$ is obvious. This is
because in the frame of the Alfv\'en phase speed, the kinetic energy
of particles is conserved. As particles are picked up by the wave,
they gain a bulk acceleration along ${\mathbf B}_0$.

\section{Pick-up of continuously created ions}

\begin{figure}[b]
\epsscale{0.95} \plotone{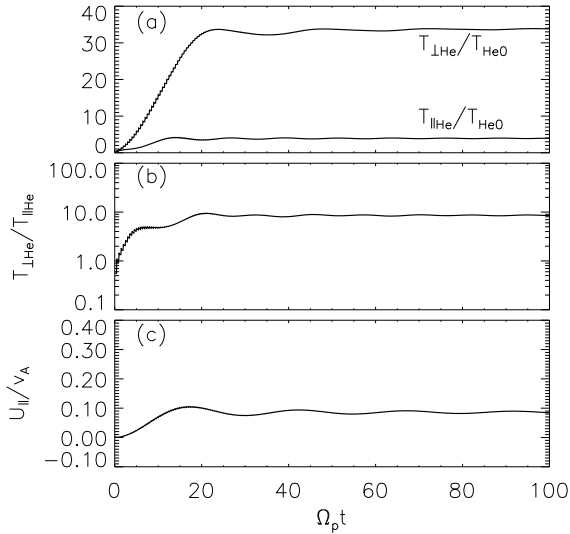} \caption{The time history of (a)
$T_{\perp He}/T_{He0}$ and $T_{\parallel He}/T_{He0}$, (b) $T_{\perp
He}/T_{\parallel He}$, and (c) $U_\parallel/v_A$. Here
$kv_A/\Omega_p=0.25\pi/125\approx 0.0063$, $\beta_{He}=0.01$,
$B^2_k/B^2_0=0.09$. The He$^{+1}$ ions are steadily and uniformly
created between $z=0$ and  $1000v_A \Omega^{-1}_p$ at $0<\Omega_p
t<100$. }
\end{figure}

 In a partially ionized plasma, the creation of new ions
 due to photoionization or collisional ionization is highly random.
Consider a region with a low plasma beta and where the amplitude of
an Alfv\'en wave is finite. When ions are created from neutrals,
they are subject to rapid gyromotions. The continuous creation of
ions represents a continuous phase shift in the ion gyrospeed. This
will naturally mix the phase of particles' gyrospeeds. To explore
the pickup process of continuously created ions,
\begin{figure*}[hbt]
\epsscale{0.84} \plotone{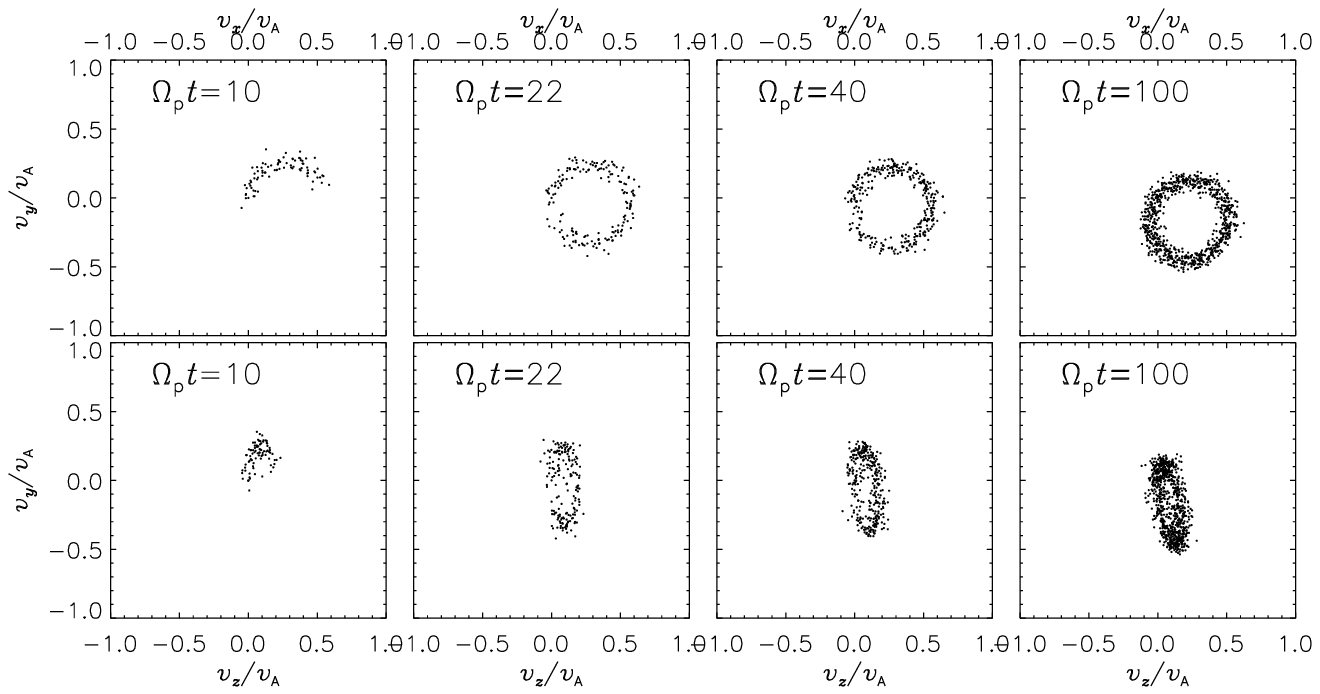} \caption{Top panel: Scatter plots
of helium He$^{+1}$ velocity in the velocity plane perpendicular to
$v_\parallel$. Bottom panel: scatter plots of helium He$^{+1}$
velocity in the plane parallel to the background magnetic field.
Relevant parameters are those in Fig.3.}
\end{figure*}
a new test particle simulation is conducted. The Alfv\'en wave and
the simulation box are the same as in section 2. However, 1000
particles are deployed every $0.25\Omega^{-1}_p$ at $0\leq \Omega_p
t \leq 100$ (one particle per cell). The beta value of the newly
created He$^{+1}$ ions is 0.01 and they have zero flow speed. The
results of the simulation are shown in Figs. 3 and 4, which are
plotted in the same way as Figs. 1 and 2.  Note that in the $v_x -
v_y$ plane, because the radius of the ring (Fig.3) is $B_kv_A/B_0$,
the origin, where new particles are created around there, is always
on the ring. When new He$^{+1}$ ions are continuously created, the
heating and scattering process are completed when they have filled
the ring of radius $B_kv_A/B_0$ in the $v_x - v_y$ plane. This only
takes one gyroperiod of He$^{+1}$ ions, or when $\Omega_pt=8\pi$, as
shown in Fig.3 and 4. Hence the heating of these particles is
extremely rapid. The ion creation process was terminated at time
$t=100\Omega^{-1}_p$, however no change is observed as $\Omega_pt
\geq 8\pi$. We believe that the rapid heating due to Alfv\'en waves
reported by Wang et al. (2006) is due to their deployment of
particles within one gyroperiod, even though they used many wave
modes in their computations.

\section{Discussion for potential applications}

The Alfv\'en wave pickup of newly created ions may have applications
at the top of the chromosphere or lower transition region.
 Let's consider the region with an ion cyclotron
frequency much higher than Coulomb collision frequencies, low plasma
beta and finite amplitude Alfv\'en waves. For neutral helium, when
wave frequencies are higher than the ion neutral collision
frequencies, the perturbed ion and neutral velocity will have a
considerable difference between their phases and amplitudes (De
Pontieu \& Haerendel 1998). In coronal funnels, the plasma beta at
the lower transition region is much smaller than unity (Li 2003).
For finite amplitude waves, the pickup process described above may
operate and heat newly created He$^{+1}$ ions. Due to collisions,
the temperature is not expected to reach the value given by
(\ref{Q13}). Instead, the energy will be transferred to protons and
electrons via Coulomb coupling. Obviously, the thermal energy of
He$^{+1}$ gained through the pickup process will be passed to
He$^{+2}$ later on when He$^{+2}$ are created. As the ionization
continues to the transition region (Hansteen et al. 1997), the
pickup process will contribute to the plasma heating there.

The pickup of instantly created ions by an Alfv\'en wave or the
pickup of ions by an Alfv\'en wave instantly introduced into a
plasma is quite efficient even for low frequency waves in
collisionless plasmas. The pickup process is complete when
$kv_{th}t=\pi$ or $t={v_A \over 2 v_{th}}t_A$, where $t_A$ is the
period of the wave. Transverse oscillations of active region coronal
loops that suddenly appear with periods of 2$-$33 minutes have been
frequently observed by TRACE (Nakariakov et al. 1999; Aschwanden et
al. 2002). The oscillations usually just last several wave periods.
The relative amplitude $B_k/B_0$ of these waves can reach 0.05
(Aschwanden et al. 2002). For typically observed 10$^6$ K coronal
loop with density of $10^9$ cm$^{-3}$ and magnetic field of 30 G,
the plasma beta value is 0.00385. Due to the pickup process, the
perpendicular temperature of helium ions and oxygen ions may be
increased to 3.6 and 11.4 times of their original values,
respectively.

\begin{acknowledgements}
   This work is supported by a PPARC Rolling Grant to University
     of Wales Aberystwyth and by the National Science Foundation of China (NSFC)
     under grants 40336052, 40674093.
\end{acknowledgements}


\begin{thebibliography}{}

\bibitem[]{416}Aschwanden, M.J., De Pontieu, B., Schrijver, C.K., \&
Title, A.M., 2002, Sol Phys., 206, 99.

\bibitem[]{419}De Pontieu, B., and Haerendel, G., 1998, A\&A, 338, 729.

\bibitem[]{422}De Pontieu, B., Martens, P.C.H, and Hudson, H.S., 2001,
ApJ, 558, 859.

\bibitem[]{425}Hansteen, V. H., Leer, E., \& Holzer, T. E. 1997, ApJ, 482, 498

\bibitem[]{427}Hollweg, J. V. 1986, J. Geophys. Res., 91, 4111

\bibitem[]{429}Leake, J.E., Arber, T.D., and Khodachenko, M.L., 2005,
A\&A, 442, 1091.

\bibitem[]{432}Li, X., 2003, A\&A, 406, 345.

\bibitem[]{434}Li, X., and Habbal, S.R., 2003, ApJ, 598, L125.

\bibitem[]{436}Lu, Q. M., and Wang, S., 2005, Geophys. Res. Lett., 32,
L03111.

\bibitem[]{439}Lu, Q. M., Wu, C. S., and Wang, S., 2006, ApJ, 638, 1169

\bibitem[]{441}Nakariakov, V., Ofman, L., Deluca, E.E., Roberts, N., Davilla, J.M., 1999,
Science, 285, 862.

\bibitem[]{444}Smith, E.J., Balogh A., Neugebauer, M., \& McComas, D.,
1995, Geophys. Res. Lett., 22, 3381.

\bibitem[]{447}Ulrich, R.K., 1996, ApJ, 465, 436.

\bibitem[]{449}Wang, C. B., Wu, C. S., \& Yoon, P. H., 2006, Phys. Rev. Lett., 96,
125001

\bibitem[]{452}Wu, C. S., Yoon, P. H., \& Chao, J. K., 1997, Phys. Plasmas, 4, 856
\end{thebibliography}
\end{document}